\documentclass[lettersize,journal,hidelinks]{IEEEtran}
\usepackage{amsmath,amssymb,amsfonts}
\usepackage{algorithmic}
\usepackage{algorithm}
\usepackage{array}
\usepackage{textcomp}
\usepackage{stfloats}
\usepackage{url}
\usepackage{verbatim}
\usepackage{graphicx}
\usepackage{cite}
\usepackage{lipsum}
\usepackage[acronym,shortcuts]{glossaries}
\usepackage[dvipsnames]{xcolor}
\usepackage{bm}
\usepackage{caption}
\usepackage{subcaption}
\usepackage{relsize}
\usepackage{soul}
\usepackage{outlines}
\usepackage{isomath}
\usepackage{orcidlink}
\usepackage[bottom]{footmisc}
\usepackage{comment}
\usepackage{mleftright}
\usepackage{orcidlink}

\renewenvironment{IEEEbiography}[1]
  {\IEEEbiographynophoto{#1}}
  {\endIEEEbiographynophoto}

\newacronym{V2X}{V2X}{vehicle-to-everything}
\newacronym{OS-QSM}{OS-QSM}{optimised scalable QSM}
\newacronym{GSM}{GSM}{generalised spatial modulation}
\newacronym{FDFR}{FDFR}{full-diversity full-rate}
\newacronym{mMIMO}{mMIMO}{massive multiple-input multiple-output}
\newacronym{MIMO}{MIMO}{multiple-input multiple-output}
\newacronym{MU}{MU}{multi-user}
\newacronym{OFDM}{OFDM}{orthogonal frequency-domain multiplexing}
\newacronym{IM}{IM}{index modulation}
\newacronym{IoT}{IoT}{Internet-of-Things}
\newacronym{QSM}{QSM}{quadrature spatial modulation}
\newacronym{BP}{BP}{belief propagation}
\newacronym{GaBP}{GaBP}{Gaussian belief propagation}
\newacronym{SM}{SM}{spatial modulation}
\newacronym{IQ}{IQ}{in-phase and quadrature}
\newacronym{ML}{ML}{machine learning}
\newacronym{BER}{BER}{bit error rate}
\newacronym{P2P}{P2P}{point-to-point}
\newacronym{AWGN}{AWGN}{additive white Gaussian noise}
\newacronym{SWIPT}{SWIPT}{simultaneous wireless information and power transfer}
\newacronym{CSI}{CSI}{channel state information}
\newacronym{MQAM}{$M$-QAM}{$M$-ary quadrature amplitude modulation}
\newacronym{IC}{IC}{interference cancellation}
\newacronym{SGA}{SGA}{scalar Gaussian approximation}
\newacronym{CLT}{CLT}{central limit theorem}
\newacronym{PDF}{PDF}{probability density function}
\newacronym{GB-ISTA}{GB-ISTA}{greedy boxed iterative soft-thresholding algorithm}
\newacronym{MP}{MP}{message passing}
\newacronym{WL}{WL}{wireless localization}
\newacronym{SD}{SD}{sphere decoder}
\newacronym{FD}{FD}{full-duplex}
\newacronym{STC}{STC}{space-time coding}
\newacronym{SotA}{SotA}{state-of-the-art}
\newacronym{IER}{IER}{index vector error rate}
\newacronym{B5G}{B5G}{beyond fifth generation}
\newacronym{STC-SM}{STC-SM}{space-time coded SM}
\newacronym{STC-QSM}{STC-QSM}{space-time coded QSM}
\newacronym{SC-IM}{SC-IM}{single-carrier IM}
\newacronym{SSK}{SSK}{space shift keying}
\newacronym{mmWave}{mmWave}{millimeter-wave}
\newacronym{THz}{THz}{Terahertz}
\newacronym{RIS}{RIS}{reflective intelligence surface}
\newacronym{RF}{RF}{radio frequency}
\newacronym{STBC}{STBC}{space-time block code}
\newacronym{MMSE}{MMSE}{minimum mean-squared-error}
\newacronym{CS}{CS}{compressive sensing}
\newacronym{i.i.d.}{i.i.d.}{independent and identically distributed}
\newacronym{JCAS}{JCAS}{joint communication and sensing}
\newacronym{ISAC}{ISAC}{integrated sensing and communication}
\newacronym{JRC}{JRC}{joint radar-communications}
\newacronym{SE}{SE}{spectral efficiency}
\newacronym{EE}{EE}{energy efficiency}
\newacronym{RBL}{RBL}{rigid body localization}
\newacronym{RBT}{RBT}{rigid body tracking}
\newacronym{SC-RBL}{SC-RBL}{soft-connected RBL}
\newacronym{W-RBL}{W-RBL}{\underline{wireless} RBL}
\newacronym{GA}{GA}{genie-aided}
\newacronym{MC}{MC}{matrix completion}
\newacronym{EA}{EA}{``\emph{estimate-then-average}''}
\newacronym{AE}{AE}{``\emph{average-then-estimate}''}
\newacronym{IRS}{IRS}{intelligent reflecting surface}
\newacronym{RSSI}{RSSI}{received signal strength indicator}
\newacronym{PSO}{PSO}{particle swarm optimization}

\newacronym{NTN}{NTN}{non-terrestrial networks} 
\newacronym{6G}{6G}{sixth-generation}
\newacronym{3D}{3D}{three-dimensional}
\newacronym{D2D}{D2D}{device-to-device}
\newacronym{RR}{RR}{round-robin}
\newacronym{DA}{DA}{Dutch auction}
\newacronym{CWFL}{CWFL}{clustered WFL}
\newacronym{WFL}{WFL}{wireless federated learning}
\newacronym{RSMA}{RSMA}{rate splitting multiple access}

\newacronym{TDMA}{TDMA}{time-domain multiple access}
\newacronym{NOMA}{NOMA}{non-orthogonal multiple access}

\newacronym{CT}{CT}{compute-then-transmit}
\newacronym{SDP}{SDP}{semidefinite programming}
\newacronym{FP}{FP}{fractional programming}
\newacronym{CF-mMIMO}{CF-mMIMO}{cell free massive MIMO}
\newacronym{iid}{i.i.d.}{independent and identically distributed}
\newacronym{DL}{DL}{downlink}
\newacronym{UL}{UL}{uplink}
\newacronym{MDS}{MDS}{multidimensional scaling}

\newacronym{SIC}{SIC}{successive interference cancellation}
\newacronym{BS}{BS}{base station}
\newacronym{TX}{TX}{transmit}
\newacronym{RX}{RX}{receive}

\newacronym{SISO}{SISO}{single-input single-output}

\newacronym{SINR}{SINR}{signal-to-interference-and-noise ratio}
\newacronym{FL}{FL}{federated learning}
\newacronym{CPU}{CPU}{central processing unit}
\newacronym{KNN}{KNN}{K-nearest-neighbor}

\newacronym{GD}{GD}{gradient descent}

\newacronym{RSS}{RSS}{received signal strength}
\newacronym{FIM}{FIM}{fisher information matrix}
\newacronym{ToA}{ToA}{time of arrival}
\newacronym{AoA}{AoA}{angle of arrival}
\newacronym{ADoA}{ADoA}{angle difference of arrival}
\newacronym{GP}{GP}{Gaussian process}
\newacronym{2D}{2D}{two-dimensional}
\newacronym{GPR}{GPR}{Gaussian process regression}
\newacronym{GNSS}{GNSS}{global navigation satellite systems}

\newacronym{RRH}{RRH}{remote radio head}
\newacronym{GPS}{GPS}{Global Positioning System}
\newacronym{RFID}{RFID}{radio frequency identification}
\newacronym{TCAS}{TCAS}{traffic alert and collision avoidance systems}
\newacronym{RMSE}{RMSE}{root mean square error}
\newacronym{SGD}{SGD}{stochastic gradient descent}

\newacronym{CU}{CU}{computing unit}
\newacronym{DM-MIMO}{DM-MIMO}{distributed massive multiple-input multiple-output}
\newacronym{LOS}{LOS}{line-of-sight}
\newacronym{NLOS}{NLOS}{non-line-of-sight}
\newacronym{ROI}{ROI}{region of interest}
\newacronym{AP}{AP}{access point}
\newacronym{PoCs}{PoCs}{projections onto convex sets}
\newacronym{TDOA}{TDOA}{time difference of arrival}
\newacronym{DoA}{DoA}{direction of arrival}
\newacronym{UE}{UE}{user equipment}
\newacronym{dB}{dB}{decibel}

\newacronym{CG}{CG}{conjugate gradient}
\newacronym{SC}{SC}{soft-connected}
\newacronym{CRLB}{CRLB}{Cramér-Rao Lower Bound}
\newacronym{PoA}{PoA}{phase of arrival}
\newacronym{UAV}{UAV}{unmanned aerial vehicle}
\newacronym{VR}{VR}{virtual reality}

\newacronym{SLAM}{SLAM}{simultaneous localization and mapping}
\newacronym{WLS}{WLS}{weighted least square}
\newacronym{JSCC}{JSCC}{joint sensing communication and computing}
\newacronym{SMDS}{SMDS}{super multidimensional scaling}
\newacronym{EDM}{EDM}{euclidean distance matrix}
\newacronym{LDPC}{LDPC}{low density parity check}
\newacronym{MAP}{MAP}{maximum a posteriori}

\begin{document}

\title{ Enabling Next-Generation V2X Perception: \\ Wireless Rigid Body Localization and Tracking\\[-1ex]}

\author{Niclas~F\"uhrling\textsuperscript{\orcidlink{0000-0003-1942-8691}}, Hyeon~Seok~Rou\textsuperscript{\orcidlink{0000-0003-3483-7629}}, \IEEEmembership{Graduate Student Members, IEEE},\\
Giuseppe~Thadeu~Freitas~de~Abreu\textsuperscript{\orcidlink{0000-0002-5018-8174}}, David~Gonz{\'a}lez~G.\textsuperscript{\orcidlink{0000-0003-2090-8481}}, \IEEEmembership{Senior Members, IEEE}, and Osvaldo~Gonsa\textsuperscript{\orcidlink{0000-0001-5452-8159}}

\vspace{-6ex}

\thanks{N.~F\"uhrling, H.~S.~Rou and G.~T.~F.~Abreu are with the School of Computer Science and Engineering, Constructor University, Campus Ring 1, 28759, Bremen, Germany (emails: [nfuehrling, hrou, gabreu]@constructor.university).}
\thanks{D.~Gonz{\'a}lez~G. and O.~Gonsa are with the Wireless Communications Technologies Group, Continental Automotive Technologies GmbH, Guerickestrasse 7, 60488, Frankfurt am Main, Germany (emails: david.gonzalez.g@ieee.org, osvaldo.gonsa@continental-corporation.com).}
}

\markboth{Submitted to the IEEE Vehicular Technology Magazine Special Issue}%
{N.~F\"{u}hrling \MakeLowercase{\textit{et al.}}}


\maketitle

\begin{abstract}
\Ac{V2X} perception describes a suite of technologies used to enable vehicles to perceive their surroundings and communicate with various entities, such as other road users, infrastructure, or the network/cloud.
With the development of autonomous driving, \ac{V2X} perception is becoming increasingly relevant, as can be seen by the tremendous attention recently given to \ac{ISAC} technologies.
In this context, \ac{RBL} also emerges as one important technology which enables the estimation of not only target's positions, but also their shape and orientation.
This article discusses the need for \ac{RBL}, its benefits and opportunities, challenges and research directions, as well as its role in the standardization of the \ac{6G} and \ac{B5G} applications.
\end{abstract}

\glsresetall
\glsresetall

\vspace{-4ex}
\section{Introduction}
%
%
%
%
%

\Ac{WL} can be regarded as an early example of \ac{ISAC} that demonstrates how communication signals can also be used to extract environmental information ($i.e.$, the location of users) using signals originally indented for communications \cite{LimaAccess2021}.
These capabilities become particularly relevant in certain industry verticals, such as intelligent transportation systems (ITS) and automotive, where there is an increasing need for \ac{V2X} perception and convergence of previously independent functions, such as sensing, communication, and computing.
Indeed, \ac{WL} has been successfully used in a wide range of applications including security, smart homes, industrial automation, robotics, vehicular networks, localization of \acp{UAV}, and more.

In terms of the fundamental concept or metrics for \ac{WL}, $i.e.$, the type of information extracted from wireless signals for the purpose of localization, many \ac{SotA} techniques have been proposed, including, methods based on finger-prints, \ac{RSSI}, \ac{AoA}, radio range, mere connectivity or hop-counts, and combinations of them.
%

The \ac{WL} literature also exhibits vast diversity in terms of the underlying mathematical approaches used in the design of positioning algorithms, with examples ranging from purely algebraic methods exploiting \ac{MDS}, and its further developed \ac{SMDS} variant, methods based on optimization-theoretical techniques, such as \ac{SDP}, \ac{PoCs}, \ac{FP} and \ac{MC}, to schemes based on \ac{ML}, which gain more popularity in the recent years.

Finally, in addition to the aforementioned methods, abundance also exists in terms of offered features, such as robustness against noise, bias, \ac{LOS} and \ac{NLOS} conditions, and mitigation of effects, including not only information scarcity, uncertainty of anchor points, co-existence of near- and far-field waves, but also physical problems, such as missing links for measurements that are needed for the positioning algorithms.
These challenges will be further discussed in Section \ref{sec:chal}.

Despite the healthy breadth of topics covered by the literature, one important \textbf{practical aspect} that has not been fully addressed by most of the \ac{SotA} \ac{WL} methods (nor commercial wireless systems, $e.g.$, 5G) is the fact that, in many use cases, targets could be better represented as \ac{3D} objects.
Aiming to address this matter, a growing literature is emerging, in which each target is modeled not as a single point, but as a group of inter-connected points with a fixed and known arrangement, $i.e$, a rigid body \cite{ChepuriTSP2014,DongTWC2023,Chen_2015,Nic_RBL}.
In our view, a rigid body representation of objects would enhance most of the applications mentioned before, such as control, mobility, and safety for autonomous robots, vehicles, and \acp{UAV}.

Thanks to the inherent inclusion of target shape and orientation models, which can be either used as prior information or jointly estimated within the solution of the localization problem, these \ac{RBL} techniques, have been shown to provide higher accuracy than \textit{conventional} (point-based) \ac{WL} methods.
Thus, \ac{RBL} expand the notion of a single \textit{reference} position to position, translation, and rotation, which are very valuable information in many realistic scenarios.
In view of the later, a significant amount of work and schemes for \ac{RBL} have appeared.
%
%
Remarkably, anchorless localization methods will become more relevant in the future, taking into account the fact that sidelink-based positioning has recently been added to the 5G specifications in 3GPP release 18, as well as \cite{02:00074}, which is a positioning technique purely operating between \acp{UE}.

With the recent popularity of autonomous driving, extending the idea of rigid body localization even further leads to rigid body tracking approaches that can be used for collision detection.
%
%
%
In that sense, in \cite{Chen_2015}, a two-stage approach was used to estimate rotation, translation, angular velocity and translational velocity by range and Doppler measurements, making use of various \ac{WLS} minimizations.
\newpage

Emphasizing the distinction between \ac{WL}/\acf{ISAC} \cite{LimaAccess2021}, where wireless connectivity is a fundamental component of the technology, and other approaches based on, $e.g.$, inertial sensing 
or machine vision, which are outside of the scope of this article,
it can be said based on the above that with respect to \ac{RBL}, the current focus of the \ac{WL}/\ac{ISAC} research community is essentially extending/applying the multitude of techniques found in the vast \ac{ML} literature to the \ac{RBL} problem,  which could be jointly referred to as \ac{W-RBL}.

The rest of the article is organized as follows:
Section \ref{sec:stand}, provides industry and standardization perspectives on the need for adopting \ac{RBL} features in beyond-5G systems, $e.g.$, \ac{6G}, as well as potential use cases and applications.
Section \ref{sec:sys} offers a brief introduction of \acf{RBL}, defines the most relevant mathematical relationships, and discusses the transition from point-based localization to rigid body localization.
Next, in Section \ref{sec:chal}, main applications, challenges and opportunities that come along with the usage of rigid bodies are discussed in detail.
Finally, Section \ref{sec:conc} closes the article with final remarks and future research directions.

\vspace{-1ex}
\section{Rigid Body Localization in beyond-5G and 6G}
\label{sec:stand}

Positioning services have been standardized in existing 3GPP commercial networks, such as Long term Evolution (LTE, 4G) and New Radio (NR, 5G), mainly as a \textit{necessary complement} to conventional Global Navigation Satellite Systems (GNSS), for certain mission critical applications with stringent requirements in terms of positioning accuracy, latency, availability, and/or trustworthiness/reliability. In general terms, 5G positioning features include network-based positioning, sidelink-based positioning (without direct use of network signals), and hybrid-enhancements which leverage positioning capabilities in other available radio access technologies (RATs), such as ultra-wide band (UWB), GNSS, Bluetooth, WiFi, etc. A variety of methodologies have been deployed including time difference, round trip time (RTT), carrier phase measurements, angle of arrival/departure (AoA, AoD), etc., and improvements have been made in terms of signaling overhead, coverage scenarios, integrity, accuracy, etc. Since the beginning of 5G-Advanced (5G-A, release 18 and beyond) positioning enhancements based on Artificial Intelligence and Machine Learning (AI/ML) have also been studied and are currently being specified. However, a more general framework for RBL has not been considered yet in 5G systems, and hence, it is an interesting (and required) \textbf{positioning expansion/enhancement} to be developed for \mbox{IMT-2030}~\cite{02:00074}, $i.e.$, 6G. 

\vspace{-1ex}
\subsection{\textbf{Standardization Perspectives and Applications}}

At the time of writing, the timeline for 6G has been sketched~\cite{01:00133} and process that will shape it within the 3GPP has also been kicked off with a workshop on 6G use cases~\cite{01:00134}. The 6G study items at the technical specification group service and system aspects (TSG SA) will be approved in September 2024 based on early inputs provided by a wide range of industry sectors (verticals), mobile network operators, regional alliances, and the International Telecommunications Union (ITU). \textbf{Sensing} and \textbf{(advanced) positioning} (potentially including RBL), have been clearly identified among the strongest drivers for 6G, and both of them have been framed within the usage scenario\footnote{IMT-2030 comprises 6 usage scenarios: 3 of them are evolution from existing 5G service types (Immersive Communication, Hyper Reliable and Low-Latency Communication, and Massive Communication) and 3 of them are \textit{expansion} towards new usage scenarios, including ISAC, Ubiquitous Connectivity, and Artificial Intelligence and Communication.} ISAC. Thus, \textbf{RBL could certainly be one of the most interesting novelties in 6G}.

Sensing and advanced positioning are key to support use cases in several verticals, such as ITS, automotive, and automated/smart  factories. Complex applications requiring localization and tracking, object reconstruction, and environment perception can be benefited from RBL capabilities. 
In the context of industrial applications~\cite{01:00135}, RBL can enhance real-time high-accuracy digital twins and industrial metaverse, which in turns enable logistics, mobile robots, and factory automation. For the automotive sector, its applicability is also clear in many use cases~\cite{09:00055} requiring real time environmental modeling and/or intelligent automated driving systems, such as automatic valet parking, lane merging/maneuvering, smart intersections, protection of vulnerable road users, and automatic charging/fueling of (electric) vehicles. In addition, RBL is also very \textit{compatible} with the use of (next-generation) \mbox{\ac{V2X} infrastructure \textit{as a sensor}} in connected/digital roads~\cite{15:00020}.   

\begin{figure*}
    \centering
    \includegraphics[width=\textwidth]{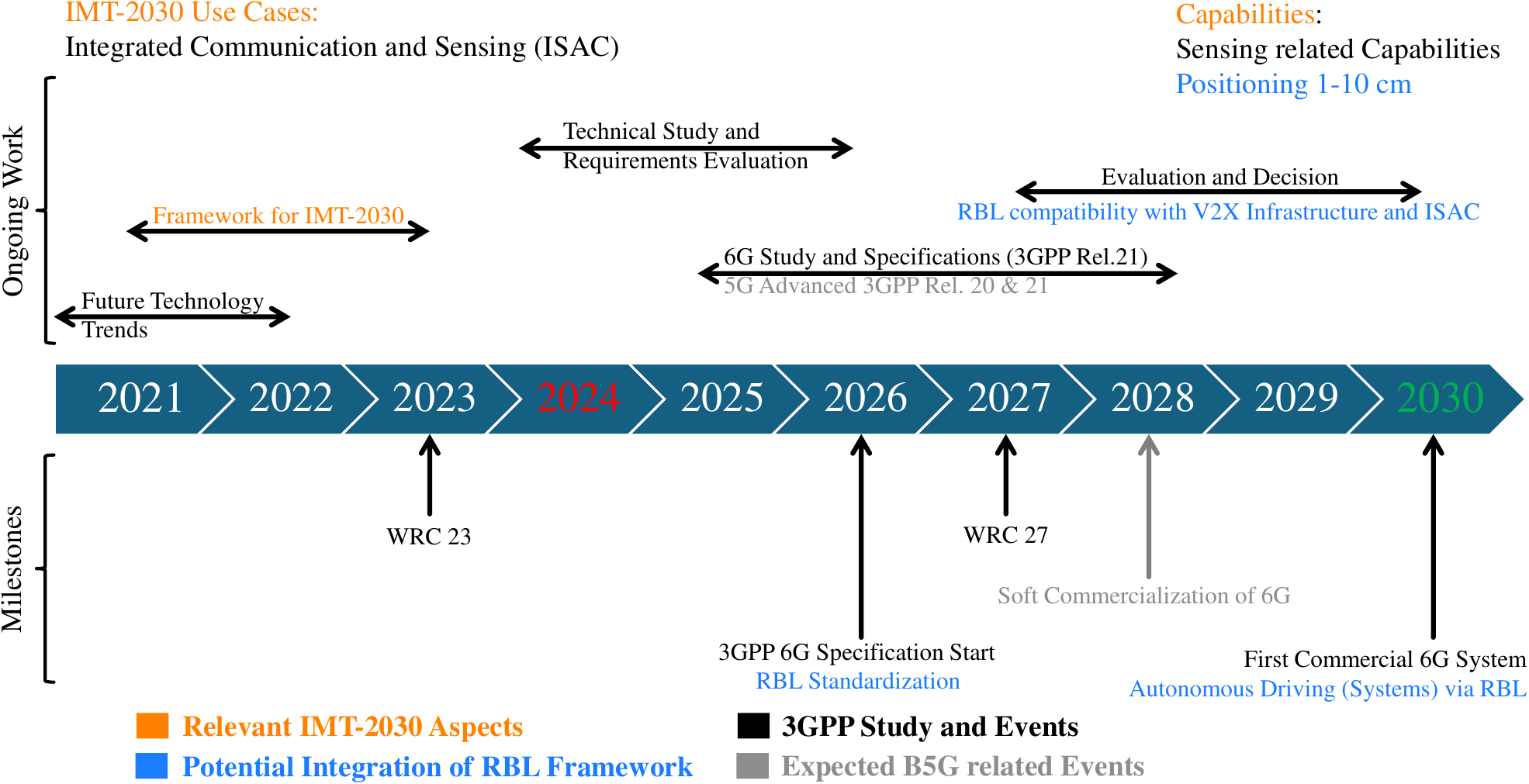}
    \caption{Suggested integration of RBL framework within the (pre)standardization of 6G \textit{in-line} with the IMT-2030 use cases and capabilities.}
    \label{fig:6G_timeline}
    \vspace{-2ex}
\end{figure*}

To emphasize the potential role of \ac{RBL} in future 6G systems, Figure \ref{fig:6G_timeline} provides an overview of past and future milestones, highlighting the capabilities and use cases specified in \mbox{IMT-2030}.

\section{Rigid Body Localization System Model}
\label{sec:sys}

\subsection{\textbf{From Wireless Point-based Localization to Wireless Rigid Body Localization (W-RBL)}}
As described before, there are many different approaches with different types of information that have been proposed for wireless localization.
In this section, we aim to describe the process of extending a standard point-based localization method to incorporate the rigid body framework in Sec. \ref{sec:rb}.
Whereas single point localization tend to use a single measurements to a set of anchors, the same can be applied to rigid bodies, where each sensor in the rigid body performs a measurement to the set of sensors, or vice versa.
The main difference lies in the fixed construction of the rigid body that need to be incorporated in the corresponding methods, having obtained information through measurements of all the individual sensors.
Thus, any localization method can effectively be reformulated to incorporate the \ac{RBL} constraints by an appropriate insertion of the \ac{RBL} framework given by equation \eqref{eq:sys_mod}, opening significant directions of improving the existing \ac{SotA} methods.

\vspace{-2ex}
\subsection{\textbf{Background of Rigid Body Localization}}
\label{sec:rb}

To gather a basic understanding of rigid bodies, we will shortly present the underlying mathematical system model of \ac{RBL}.
Given a rigid body, the position of the $k$-th node within a rigid body can be described by a \ac{2D} coordinate vector $\mathbf{c}_{k} \in \mathbb{R}^{2 \times 1}$,
%
%
which captures the $x$- and $y$-coordinates of the $k$-th node of the rigid body respectively, for all $K$ points defining the rigid body.

Next, following the \ac{SotA} system model of the standard \ac{RBL} framework \cite{ChepuriTSP2014,DongTWC2023,Chen_2015,Nic_RBL}, a change in position and orientation of a \ac{RBL} can be described by the joint affine transform of the $K$ coordinates of a rigid body, given by \vspace{-1ex}
\begin{equation}
\begin{split}
\mathbf{S} &\triangleq [\mathbf{s}_1, \cdots\!, \mathbf{s}_k, \cdots\!, \mathbf{s}_K] \\&= \mathbf{R} \!\cdot\! \overbrace{[\mathbf{c}_1, \cdots\!, \mathbf{c}_k, \cdots\!, \mathbf{c}_K]}^{\triangleq \; \mathbf{C}} +\;\! [\mathbf{t}, \cdots\!, \mathbf{t}, \cdots\!, \mathbf{t}] \\&=\mathbf{R}\mathbf{C}+ (\mathbf{t}\otimes\mathbf{1}_{1\times K}) , 
\end{split}
\label{eq:sys_mod}
\end{equation}
where $\mathbf{S} \in \mathbb{R}^{2 \times K}$ consists of the new coordinates of the rigid body nodes after the transform, $\mathbf{R}\in \mathbb{R}^{2 \times K}$ and $\mathbf{t} \in \mathbb{R}^{2 \times 1}$ respectively describe the rotation and translation equally to the $K$ original coordinates, and $\otimes$ is the kronecker product.

In addition, another important property of rigid bodies is the fixed relative Euclidean distances between the $K$ nodes in $\mathbf{C}$, which must also be satisfied in $\mathbf{S}$.
Therefore, such information is incorporated in \ac{RBL} method as a necessary constraint.

The implication for \ac{RBL} is that the known structure of the rigid body in $\mathbf{C}$ will be exploited to aid the estimation of the localized positions in $\mathbf{S}$, where the affine transform $\mathbf{R}$ and $\mathbf{t}$ is unknown, and only some noisy observation of $\mathbf{S}$, or other kinds of measurements related to $\mathbf{S}$ are available.

\begin{figure}[H]
         \centering
         \includegraphics[width=\columnwidth]{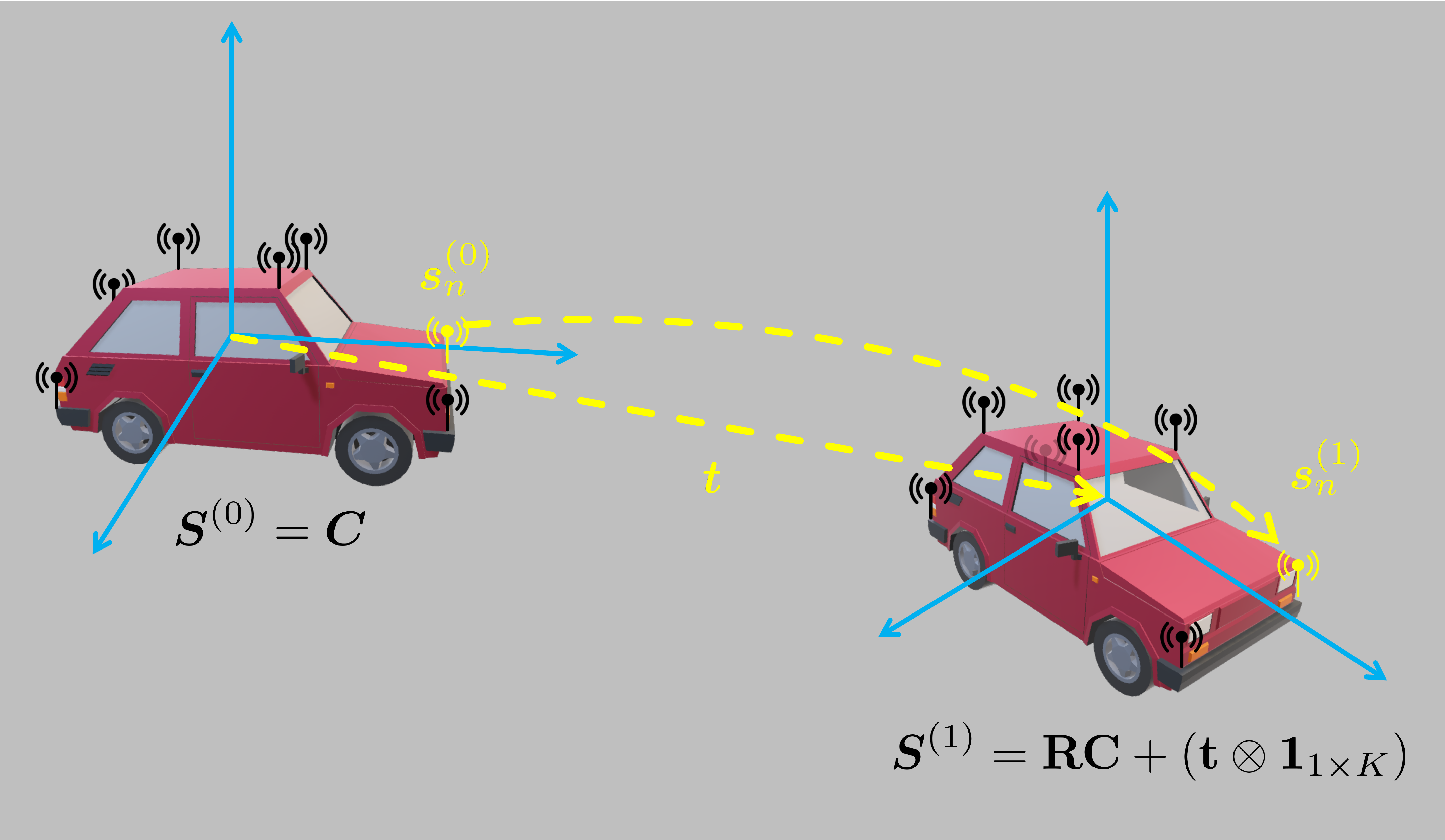}
         \caption{Illustration of a rigid body at two distinct locations $\boldsymbol{S}^{(0)}$ and $\boldsymbol{S}^{(1)}$. Without loss of generality, we set the initial to be identical to the matrix $\boldsymbol{C}$, which defines the shape and orientation of the rigid body. The second location $\boldsymbol{S}^{(1)}$ of the body relative to its initial location $\boldsymbol{S}^{(0)}$ is then determined according to equation \eqref{eq:sys_mod}, and is obtained by the transformation of $\boldsymbol{S}^{(0)}$ via a rotation matrix $\bm{R}$ and a translation vector $\boldsymbol{t}$.}
         \label{fig:RB_sys}
     \end{figure}

The above model and properties can easily be extended to the \ac{3D} scenario by adding another dimension to the coordinate vector and correspondingly in eq. \eqref{eq:sys_mod}, with out a loss of generality, where a simple illustration of a rigid body transformed by a rotation and a translation is visualized in Figure \ref{fig:RB_sys}, by vehicle turning on a street.

\vspace{-4ex}
\section{Challenges and Opportunities}
\label{sec:chal}

With the main idea of rigid bodies and a clear mathematical description defined, this chapter aims to discuss various approaches problems and solutions to \ac{RBL}, drawing an analogy to single-point localization schemes and presenting which problems have been solved for those schemes, but are open to investigate for the special case of rigid body targets.
This includes the discussion for multiple different localization approaches, the scenario where the needed observations are incomplete, thoughts about the optimal sensor deployment, as well as some insights into next-generation extensions to anchorless localization and tracking.
\newpage

\subsection{\textbf{Localization Approaches}}

\subsubsection{\textbf{Algebraic Approaches}}
\begin{figure*}[b]
    \centering
    \includegraphics[width=\textwidth]{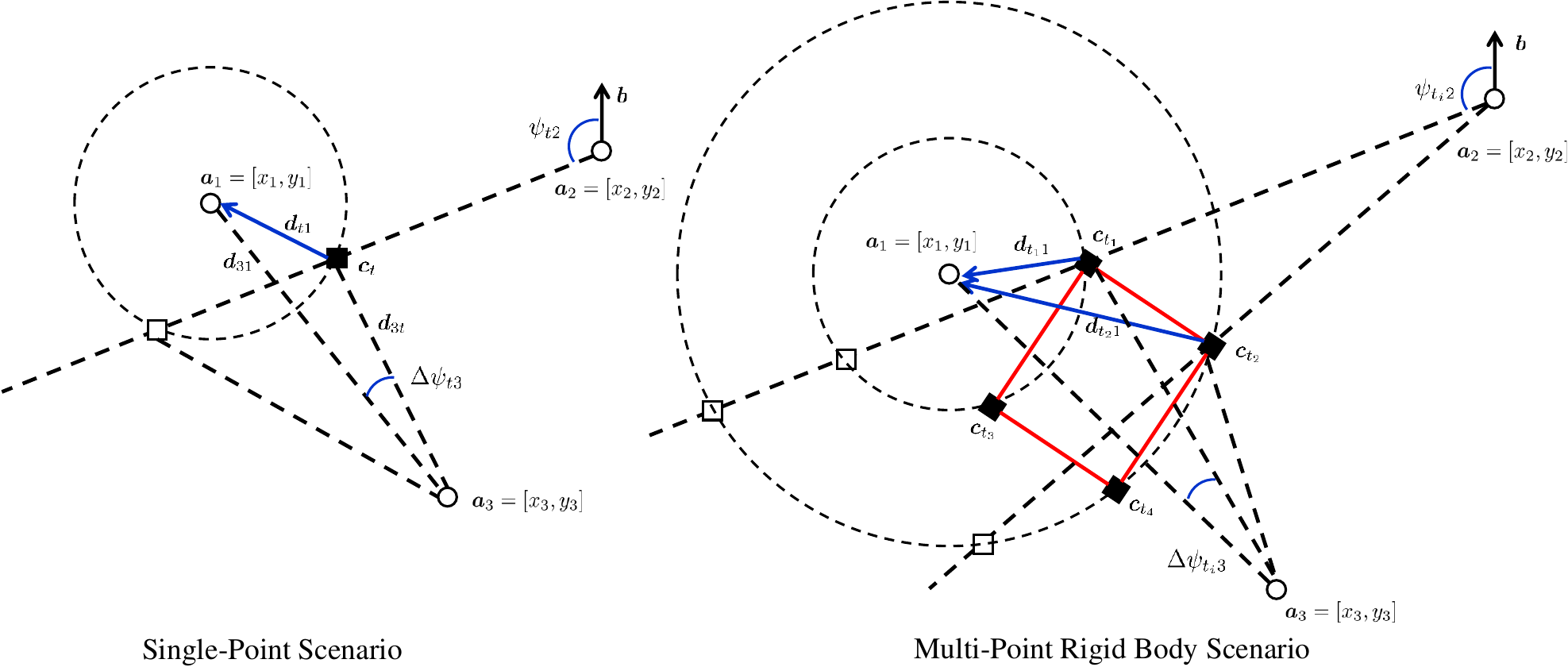}
    \caption{Combination of different measurement types for algebraic single-point localization, versus multi-point rigid body (red lines) localization, from anchors $\boldsymbol{a}_i$ (white balls) to sensor $\boldsymbol{c}_{t_i}$ (black square), for algebraic localization.}
    \label{fig:MeasComb}
    \vspace{2ex}
\end{figure*}
Inspired by the aforementioned discussion, it can be considered what type of measurements can be used and combined for localization, which measurements are the most useful, and how one can benefit from the known shape of the rigid body.
Combining all these measurements leads to different algebraic and geometric approaches resulting in diverse estimation methods, with the most well known examples being triangulation, \ac{MDS}, and \ac{SMDS} approaches \cite{Nic_RBL}.
Therefore, to give an insight on how these methods can be adapted to rigid bodies, Fig. \ref{fig:MeasComb} illustrates how different type of measurements can be combined for single-point localization, as well as \ac{RBL}.
On the left hand side, a single target is localized via range, \ac{AoA} and \ac{ADoA} measurements, where by combination of the later the position can be exactly defined.
On the right hand side, a multi-point rigid body is localized by the combination of the same measurements.
What can be observed is that due to the known shape of the rigid body one might be able to save certain measurements, since the conformation matrix adds a constraint that helps in the localization process.
Additionally it can be seen that for a rigid body, angle measurements are much more relevant than range measurements, since two points can have the same distance to an anchor, as illustrated, while the angles change.
However, which kind of measurements are the most effective and how to process them is still open to investigate.

\subsubsection{\textbf{Optimization Approaches}}
In terms of optimization, a lot of \ac{SotA} methods, can, and have been applied to rigid bodies.
Belonging to these applications, methods like least squares, \ac{SDP}, \ac{PoCs}, \ac{FP} and \ac{MC} schemes have been deployed \cite{ChepuriTSP2014,DongTWC2023}.
In contrast to the pure algebraic approaches, optimization based methods tend to describe the desired problem mathematically in a way such that through common techniques, such as gradient descent methods, an optimization problem dealing with the main objective, as well as, for example, noise constraints can be solved.
Depending on the problem statement itself, these approaches can lead to efficient estimations, possibly leading to closed-form solutions, which can also be adapted to rigid body frameworks.

\subsubsection{\textbf{Message Passing}}

The algebraic and classical optimization methods described above, while providing optimal solutions with possibly even closed-form expressions, typically require high complexity due to inherent matrix inversions and exhibit low flexibility and robustness to various additional/incompleteness of information in the system.
Alternatively, in light of the fundamental system model given in eq. \eqref{eq:sys_mod}, tailored message passing algorithms \cite{Feng_2021} may be designed to solve the \ac{RBL} problem of estimating the \ac{3D} rotation angles and translation vector.

\begin{figure}
    \centering
    \includegraphics[width=\columnwidth]{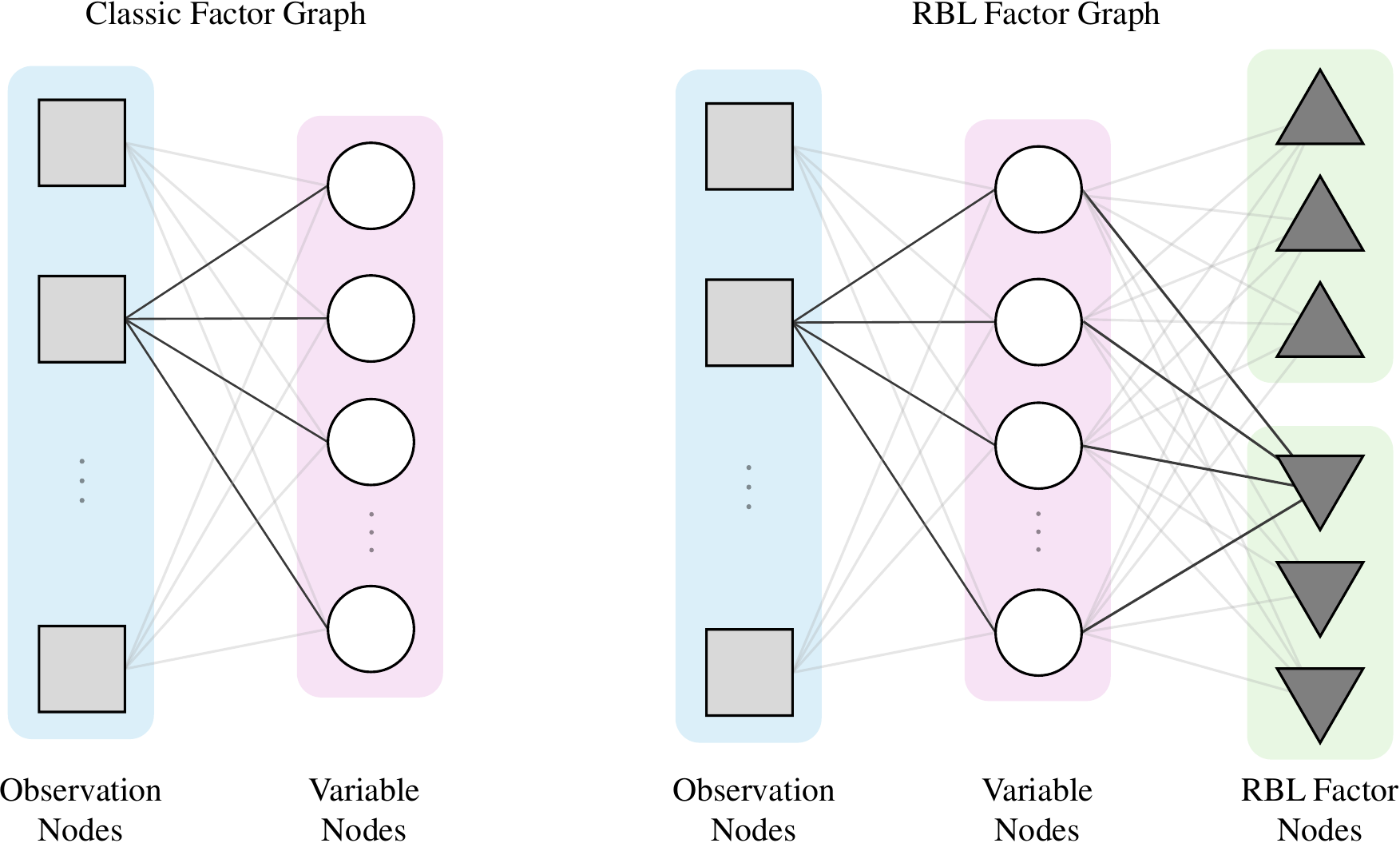}
    \caption{ An illustration of an exemplary factor graph structure incorporating the rigid body conformation (right), in comparison to a classical factor graph (left).}
    \label{fig:FG}
    \vspace{-2ex}
\end{figure}

Due to their flexibility in application and design, message passing algorithms have been thoroughly investigated and exploited in vast areas of engineering and science, including channel coding applications, data estimation, distributed machine learning and much more, and have been shown to be able to provide Bayes-optimal solutions under proper conditions.
The fundamental uniqueness of the method lies in the concept of passing locally computed messages in a factor graph structure shown in Fig. \ref{fig:FG}, where different nodes correspond to different parts of observed information and/or unknown variables of the original system, under which the key \textit{messages} -- typically the sufficient parameters of probability distributions -- are computed at each node and transmitted (\textit{passed}) over the graph edges, which represent the existence of a dependency between the nodes (\textit{i.e.,} between the observation and the variable).
In hand of the received messages, different nodes recompute and update the variable beliefs, repeating the message passing iterations until convergence.

Due to this distributed nature, the message passing algorithms are known to be robust to incomplete data and irregularity in the system model, whose framework is flexible to support the design of an estimator given the underlying system.
However, they may suffer from self-interference and divergent behavior if proper design and mitigation techniques are not applied in the formulation of the messages.

In light of the above, we highlight the message passing technique as a promising solution to the \ac{RBL} problem, which fundamentally involve observation data such as range, angle, and power measurements, whose values are related to the sensor position variables.
Furthermore, the unique relationship (constraint) between the position variables in the form of the rigid body conformation can efficiently be represented by an additional layer of edges between the variables, as exemplified in Fig. \ref{fig:FG}.
In addition, the inherent probability distribution-based operation of the message passing fits nicely with the notion of localization with multiple distributions and confidence ellipses (\textit{soft-decisions}), not to mention the robustness to unavailable observations often arising in wireless sensing scenarios, as will be elaborated in the following section.

\vspace{-2ex}
\subsection{\textbf{Incomplete Observations}}
\label{sec:MC}

\begin{figure}[t]
         \centering
         \includegraphics[width=\columnwidth]{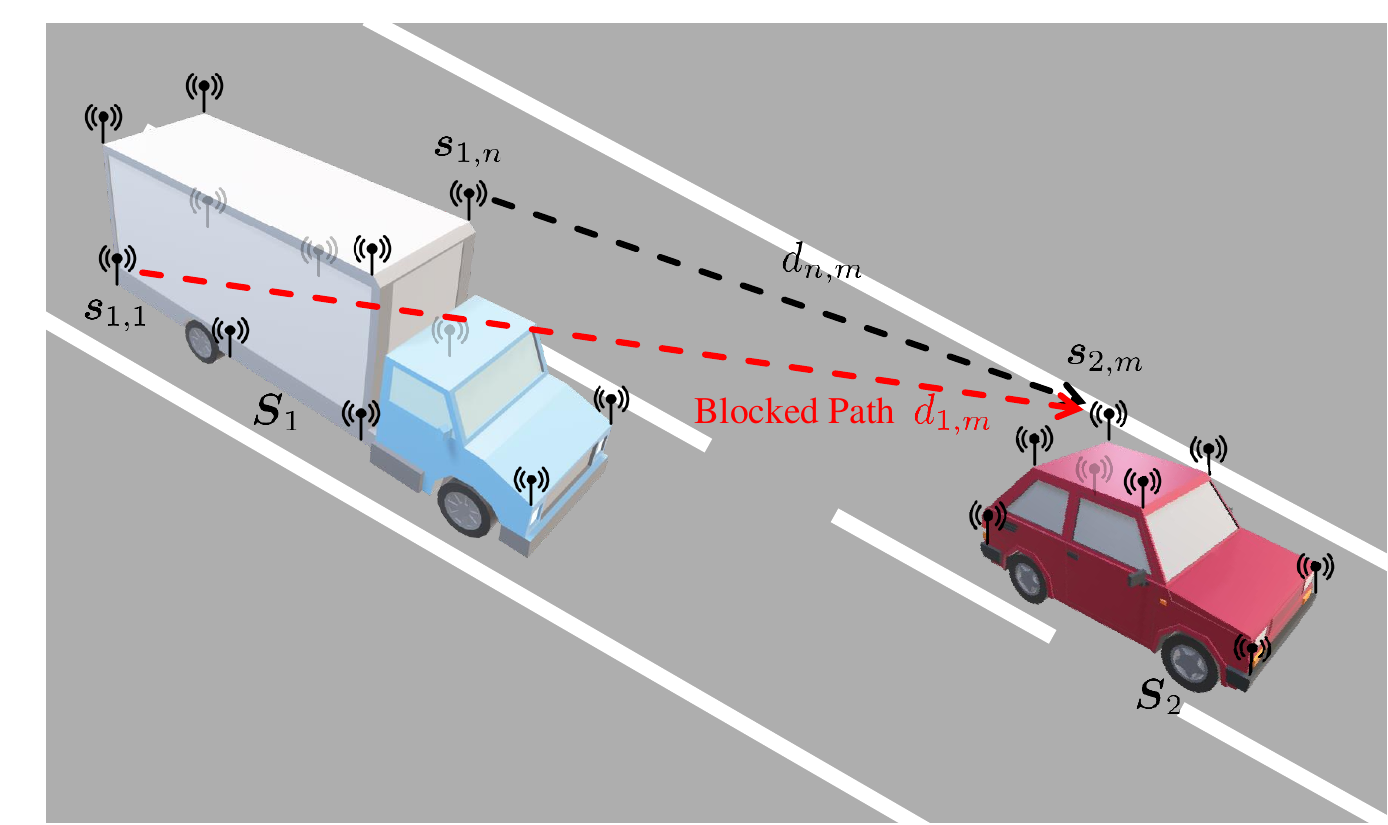}
         \caption{Illustration of two-body \ac{RBL} scenario, performing sensor to sensor range measurement. While most sensors can measure the distance, $e.g.$, $d_{n,m}$, other measurement are unavailable due to blocked paths, $e.g.$, $d_{1,m}$, depicted by the red arrow.}
         \label{fig:NLOS}
         \vspace{-2ex}
\end{figure}

As hinted by the prior sections, independent of the metric used for localization, a common issue regularly occurring is the observation of incomplete measurements, for instance due to blocked paths.
Just to name one example, when considering range measurements from anchors to sensors, a \ac{LOS} path might be blocked due to the structure of the rigid body, which leads to zeros occurring in the observed distance matrix, shown in Figure \ref{fig:NLOS}.
In the case of range measurements these zeros are especially harming, since putting a zero is equivalent to stating that the measured distance is exactly zero, which poisons the algorithms.
Therefore, these zeros have to be handled with a lot of care, since either the algorithm has to be adjusted to be aware of the missing information, or something has to be done to reconstruct the lost data.

One possible solution to this problem, which highly benefits from the structure of the rigid body, captured in the conformation matrix $\bm{C}$ are \acf{MC} solutions.
\ac{MC} literature \cite{Chen_2022} offers a wide range of different methods that can be used for completion, often used in scenarios such as recommender systems in computer science, wave channel estimation in wireless communication, but also localization algorithms in signal processing.
By knowing the conformation structure of the rigid body itself, as well as the positions of the anchors, it is possible to construct a hollow squared \acf{EDM}, containing the intra-distances of the anchors to one another, the intra-distances of the points in the rigid body itself, as well as the incomplete measured cross-distances.
This information alone is sufficient to complete the matrix with high accuracy to improve the performance significantly.
Recent contributions have proposed a discrete-aware variation that makes use of the knowledge that the values to complete can be chosen from a finite discrete alphabet set.
In theory this can be implemented for a rigid body as well, where after measuring
the incomplete distance matrix, the discrete sets can be found by taking into account all possible rotations of the rigid body, only being possible because the structure of the body is fixed and known.
Even though the discrete-aware \ac{MC} algorithm can be performed once the sets are found, the construction of the sets itself becomes a combinatorial problem for larger rigid bodies, where more research is required to find suitable algorithms for the constructing of the set.
%

\vspace{-2ex}

\subsection{\textbf{Optimal Performance and System Setup}}

Another important aspect in terms of real life scenarios and industrial standards is the sensor deployment on the rigid body itself, $i.e.$ how many sensors to use and where to put them on a specific vehicle.
Since the effect has not been studied sufficiently in the \ac{SotA}, future work on this topic needs to be done, with a precise mathematical analysis of the effects.

\begin{figure}[t]
    \centering
    \includegraphics[width=\columnwidth]{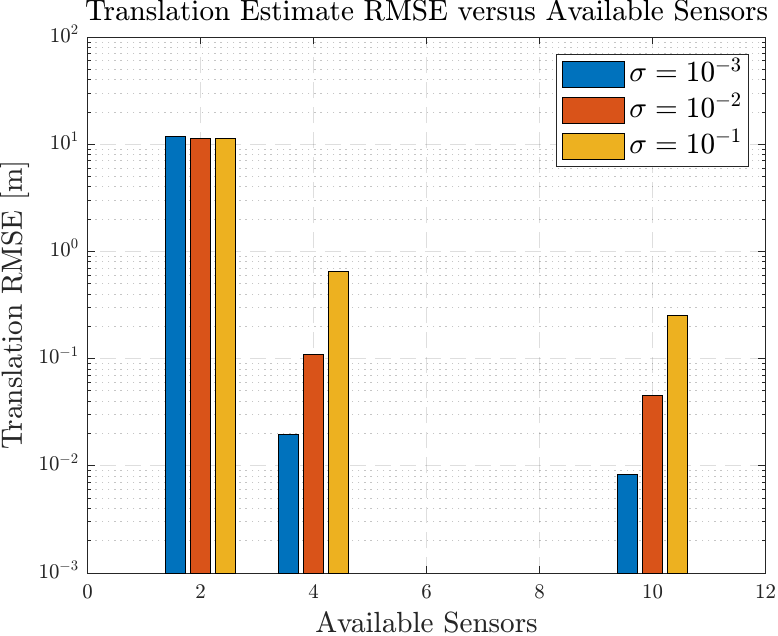}
    \caption{Translation estimate RMSE versus number of available sensors, for varying noise levels, in 3D rigid body scenario, applying the least squares solution from \cite{Chen_2015}.}
    \label{fig:Sens}
\vspace{1ex}
\includegraphics[width=\columnwidth]{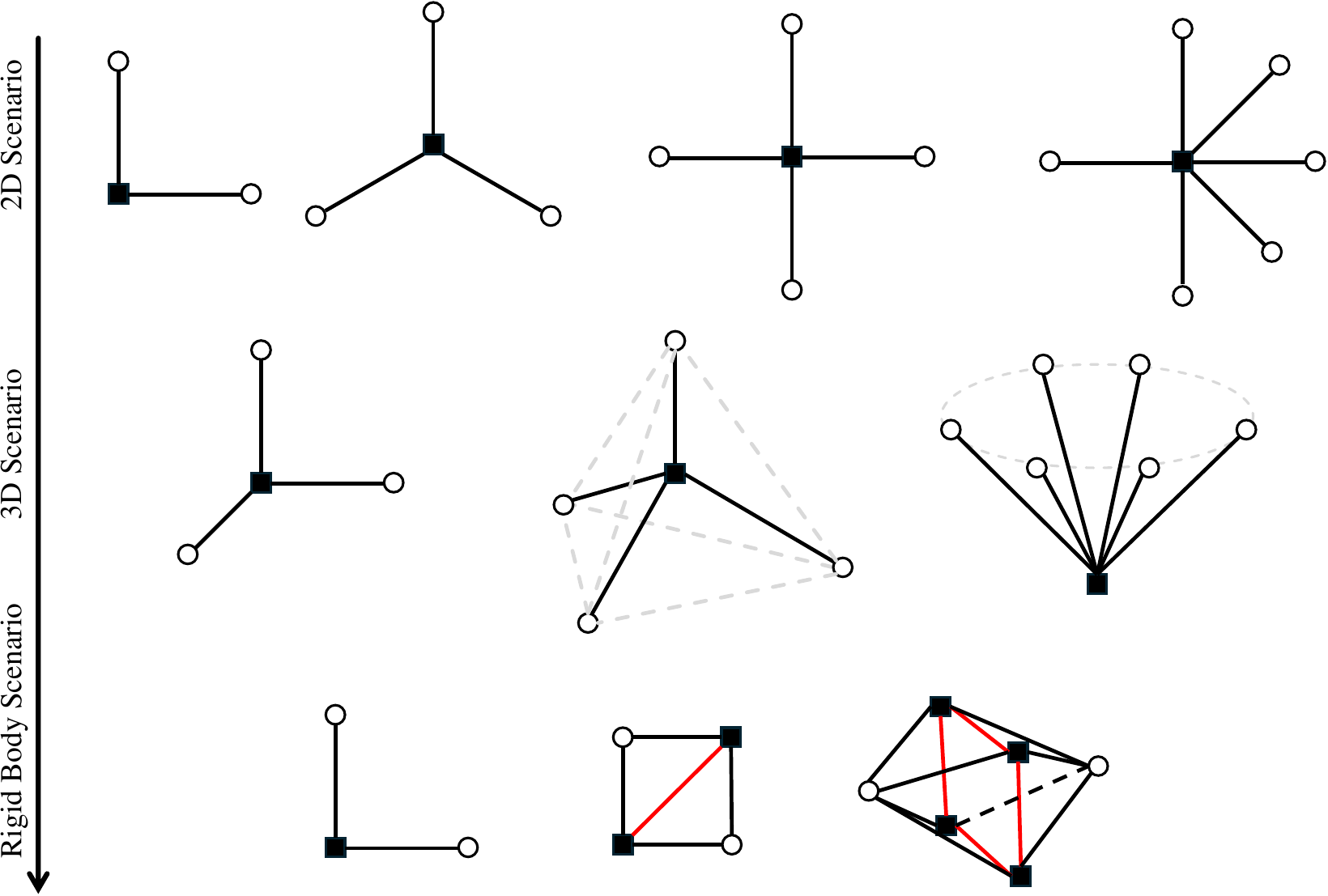}
\caption{Selection of tight frames in 2D and 3D scenarios, with circles indicating anchor positions and squares representing targets, extended to a rigid body scenario. Without lo of generality black lines indicate the connection between sensor and target, dotted-gray lines indicate the orientation in 3D and red lines indicate the conformation of the rigid body.}
\label{fig:Frames2}
\vspace{-2ex}
\end{figure}

For illustration, a simple set of simulations have been performed with a \ac{3D} vehicle that was localized via a conventional least squares minimization problem \cite{Chen_2015}, with the amount of sensors and their position varying.
Results of this can be found in Figure \ref{fig:Sens}, illustrating the translation estimate \ac{RMSE} for a varying number of sensors with different noise levels.
It can be observed that for two sensors only, no satisfying result can be obtained, where four sensors are sufficient to locate the \ac{3D} rigid body, with a large performance gain.
Furthermore, when the number of sensors increases to ten, the error decreases and starts to converge.

In terms of positioning the sensors on the rigid bodies the specific used case has to be defined first before, deploying the sensors.
As an example, take the scenario presented in Section \ref{sec:MC}, where some measurements might be missing due to \ac{NLOS} paths.
In these type of scenarios on has to make sure that the sensors are positioned in such a way that the amount of possible \ac{NLOS} paths is minimized, depending on the scenario.
On the one hand, if the sensors will connect to anchors positioned high up in road side units it might be useful to place the sensors on top of the rigid body.
On the other hand, if the rigid body is connecting to other rigid bodies, it is advisable to place them in the front and the back, such that they can connect to other rigid bodies in all possible directions.
However, for conventional single-point localization scenarios, multiple \ac{SotA} methods have investigated the problem of optimal sensors and positioning.
To name one example, \cite{zhao2013optimal} investigates the optimal anchor positioning using frame theory.
To summarize, it was found that for single targets the anchors can placed to minimize the frame potential, $i.e.$ how closely a set of anchors in a Hilbert space approximates an orthonormal basis.
For multiple targets, the anchors can be placed sequentially optimizing the the position of the anchor with respect to each of the targets position, which is build on the problem of single-target localization using \ac{AoA} measurements.
As a result, \cite{zhao2013optimal} proposed the optimal anchor placements in 2D and 3D for different amount of anchors, forming tight frames that were proved to be optimal for single targets.
However, to translate the idea on rigid bodies, Fig. \ref{fig:Frames2} first presents optimal placements for \ac{2D} and \ac{3D} scenarios, before extending the idea, illustrating difficulty of placing two sensors only for rigid bodies of higher dimensions, where for a 2D rigid body the anchors need to be placed in 3D space to be optimal, however, for a 3D object this is not possible anymore.
With the work that has been done, it is left to do further research and investigations on how the concepts can be applied to \ac{RBL}, keeping in mind that the frame theory presented only applies to the anchors surrounding the rigid bodies, put not the placement of sensors on the rigid body itself.
Similarly the effect of sensor uncertainty of the anchors, as well as the sensors on the rigid body itself need to be studied in more depth.

\vspace{-2ex}
\subsection{\textbf{Towards Next-Gen Rigid Body Localization and Tracking}}

\subsubsection{\textbf{Anchorless Multi-Rigid Body Localization}}

\begin{figure}[H]
         \centering
         \includegraphics[width=\columnwidth]{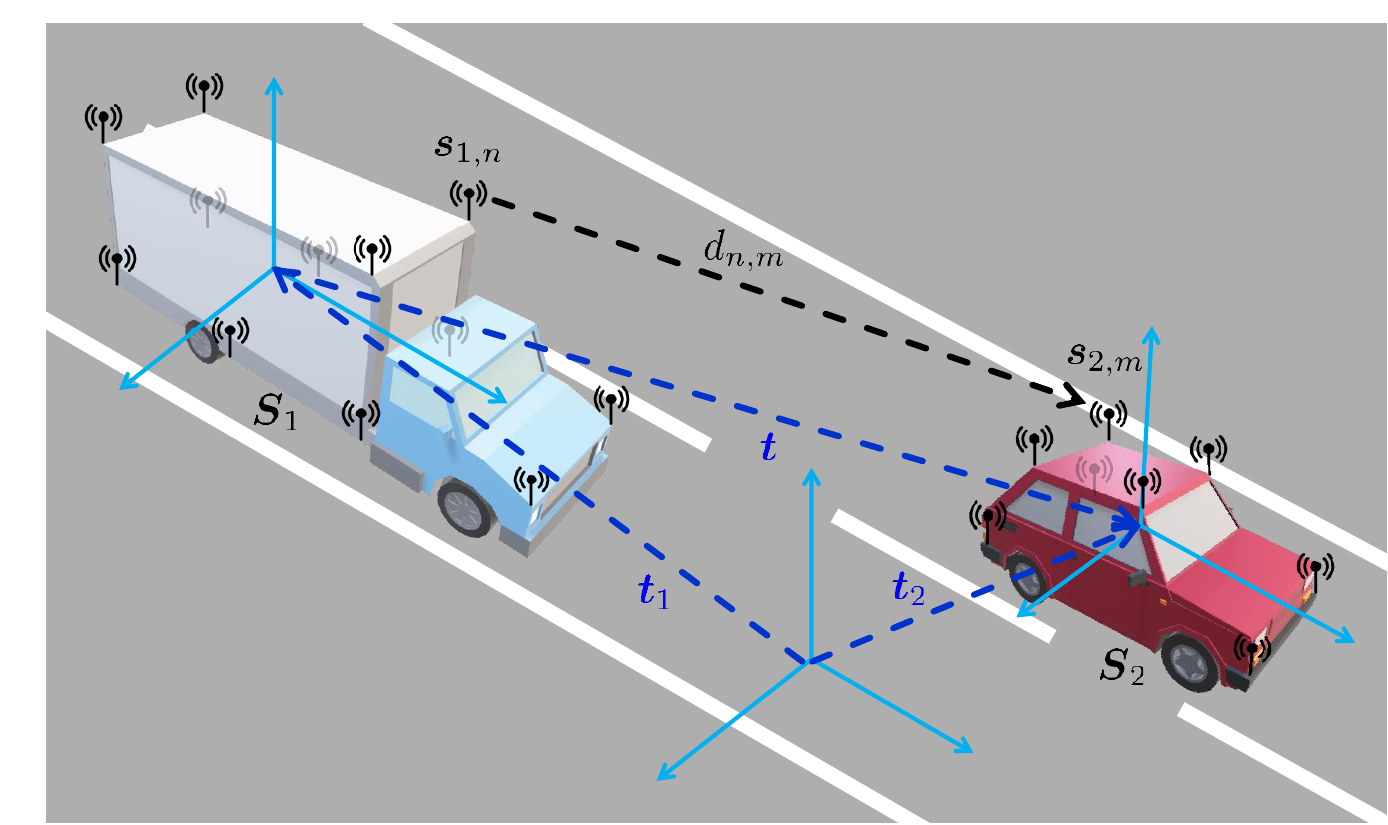}
         \caption{Illustration of two-body \ac{RBL} scenario. Each rigid body has a different shape, defined by distinct conformation matrices $\boldsymbol{C}_1$ and $\boldsymbol{C}_2$, respectively. The translation vector $\boldsymbol{t}$ between the bodies, depicted in yellow, is defined by the difference between the geometric centers of the two bodies.}
         \label{fig:V2V}
     \end{figure}

Another more recent relevant approach in terms of autonomous vehicles proposes a relative multi-rigid body localization method in an anchorless scenario, estimating the relative translation and rotation between two rigid bodies by measuring the cross-body \ac{LOS} distances between the bodies, which is visualized in Fig. \ref{fig:V2V}.
This relative multi-rigid body localization is especially useful in scenarios where there are no, or only a few anchors deployed, such that the bodies can find the relative translation and rotation even without any external sensors, which can be useful for collision detection for vehicles or \acp{UAV} or for other autonomous driving scenarios, only to name a few possible use-cases.

However, there exists one \textbf{major flaw} related to a real life implementation of such a system, which is the following.
As explained in \eqref{eq:sys_mod} and Section \ref{sec:sys}, each rigid body has a known conformation matrix $\boldsymbol{C}$.
The problem occurs when localization methods need the conformation matrices of all vehicles in order to work, which is not the case in real life, since a body can indeed measure the distances to other sensors, but does not necessarily know the other bodies structure.
To solve such issues for real life applications regulations on how knowledge on $\boldsymbol{C}_i$ should be acquired and distributed would be required.
Otherwise, egoistic methods need to be developed, which are able to perform a localization purely by measurements, without any prior knowledge about the rigid body itself, such as the one proposed in \cite{Nic_RBL}.

Nevertheless, common \ac{SotA} methods, such as \ac{MDS} can be used to obtain this knowledge and solve the problem, whereas other problems, such as missing measurements due to blocked paths can be solved by methods as described in Section \ref{sec:MC}.

\vspace{2ex}
\subsubsection{\textbf{From Rigid Body Localization to Rigid Body Tracking}}

Inspired by the aforementioned work and as a logical next step towards autonomous driving, one has to extend the concept from rigid body localization to rigid body tracking.
While for single-point based localization methods it is more intuitive to connect the model to time, adding a time constant to the rigid body system model can become a more complicated task.
Single-point based systems can easily incorporate time by applying a velocity to their point, which is not directly possible  for rigid bodies.
In terms of rigid body tracking, a lot of research has been done in the field of visual tracking, for example tracking robots, faces or general gestures by cameras.
In that sense many approaches, such as feature based tracking, edge based tracking, or sensor fusion has been proposed, which performs very well for its applications.
Nevertheless, as described before, we are focusing on \acf{W-RBL} and therefore have to search for a different approach.
What can be done to model the motion of the entire body instead of visually capturing its motion is to add an angular velocity and a translational velocity \cite{Chen_2015}.
Mathematically this can be described in simple form per sensor as

\begin{equation}
    \dot{\boldsymbol{s}}_i=[\boldsymbol{\omega}]^\times\boldsymbol{R}\boldsymbol{c}_i+\dot{\boldsymbol{t}}_i, 
    \label{eq:track}
\end{equation}
where $\dot{\boldsymbol{s}}_i$ denotes the velocity of the individual sensor in the inertial frame, $[\boldsymbol{\omega}]^\times$ being the cross product operator matrix applied to the angular velocity and $\dot{\boldsymbol{t}}_i$ being the translational velocity.

Another important aspect to keep in mind is what type of tracking to pursuit, since multiple approaches are known and can be called tracking.
The most common tracking approach is to a notion of time in the system model, which coherently influencing the velocity or similar parameters.
These type of tracking applications are commonly solved via \ac{SotA} methods, such as Kalman filters.
Another more relevant type of tracking with respect to rigid body applications is to look at moving rigid bodies in time from frame to frame, only working at one time instance at a time, such as shown in equation \eqref{eq:track}.
This type of tracking can be used especially effective in frameworks such as platooning problems.

\section{Conclusion and Outlook}
\label{sec:conc}

In this work we have provided a detailed discussion about the usage of \acf{RBL}, which is a highly promising technique for future \ac{V2X} perception and autonomous driving applications.
In that sense, we have presented the general system model of a rigid body framework, with the transition from a single point localization to \ac{RBL}, enabling the estimation of the orientation of the target instead of only the position.
Furthermore, challenges and opportunities in a vast array of fields and applications have been discussed.
The discussion about incomplete measurements directly connects to the sensor deployment part, which has shown how the amount of anchors and their placement in general can influence the performance, leading to conditions where incomplete measurements occur which can be addressed by different methods, such as matrix completion.
Further, it was discussed how modern signal processing methods, including message passing can be redesigned to fit the \ac{RBL} problem, resulting in significant complexity reductions/gains.

Even more relevant for future applications, especially for autonomous driving in rural areas where not many anchors are available, we discussed how multiple rigid bodies can perform localization in anchorless scenarios with respect to each other
and how the whole framework of \ac{RBL} can be extended to rigid body tracking.
With this in mind, future work and applications might develop scenarios where two rigid bodies belong to one object, which have a soft connection joint between them \cite{führling2023softconnectedrigidbodylocalization}, for instance in the case of large busses or trains with multiple movable sections.

Finally, as discussed in Section \ref{sec:stand} and illustrated in Figure \ref{fig:6G_timeline}, positioning applications have been standardized in all existing 3GPP commercial networks, where advanced positioning has been set as a goal for 6G, which offers great potential for future \ac{RBL} applications. 


\bibliographystyle{IEEEtran}

\vspace{-15ex}

\begin{IEEEbiography}{Niclas Führling}
   received the B.Sc. degree in electrical and computer engineering from Jacobs University Bremen, Bremen, Germany in 2022. He is currently pursuing the M.Sc. degree in electrical engineering with the University of Bremen, with a focus on communication and information technology, while working on a research project at Constructor University, focusing on 6G connectivity. His current research interests are wireless communications, signal processing, and positioning.
\end{IEEEbiography}
\vspace{-20ex}
\begin{IEEEbiography}{Hyeon Seok Rou}
 (S'19) has received the B.Sc. degree in Electrical and Computer Engineering from Jacobs University Bremen (currently Constructor University), Germany, in 2021 - where he is currently pursuing a Ph.D. in Electrical Engineering, funded by a research project from the Wireless Communications Technologies group at Continental A.G. His research interests lie in the fields of joint communications and sensing (JCAS), Bayesian statistics, multi-dimensional modulation schemes, and mmWave/sub-THz MIMO wireless communications.
\end{IEEEbiography}
\vspace{-20ex}
\begin{IEEEbiography}{Giuseppe Thadeu Freitas de Abreu}
(Senior Member, IEEE) is a Full Professor of Electrical Engineering at Constructor University, Bremen, Germany. His research interests include communications theory, estimation theory, statistical modeling, wireless localization, cognitive radio, wireless security, MIMO systems, ultrawideband and millimeter wave communications, full-duplex and cognitive radio, compressive sensing, random networks, connected vehicles networks, and many other topics. He currently serves as an editor to the IEEE Signal Processing Letters and the IEEE Communications Letters.
\end{IEEEbiography}
\vspace{-20ex}
\begin{IEEEbiography}{David González G.}
(S’06–M’15–SM’18) received a Ph.D. degree in Signal Theory and Communications from the Universitat Politècnica de Catalunya, Spain. Since 2018, David is with Continental AG, Germany, focusing on diverse aspects of wireless communications, vehicular communications (V2X), and integrated sensing and communications (ISAC). David also serves as delegate in the 3GPP for 5G standardization, ETSI, and 5GAA.
\end{IEEEbiography}
\vspace{-20ex}
\begin{IEEEbiography}{Osvaldo Gonsa}
received the Ph.D. degree in electrical and computer engineering from Yokohama National University, Japan, in 1999, and the M.B.A. degree from the Kempten School of Business, Germany, in 2012. He is currently the Head of the Wireless Communications Technologies Group, Continental AG, Frankfurt, Germany. And since 2020 also serves as a member for the GSMA Advisory Board for automotive and the 6GKom Project of the German Federal Ministry of Education and Research.
  \end{IEEEbiography}

\end{document}